# Pressure induced superconductivity in $WB_2$ and $ReB_2$ through modifying the B layers

Cuiying Pei[1#], Jianfeng Zhang[2#], Chunsheng Gong[2#], Qi Wang[1,3], Lingling Gao[1], Yi Zhao[1], Shangjie Tian[2], Weizheng Cao[1], Changhua Li[1], Zhong-Yi Lu[2], Hechang Lei[2*], Kai Liu[2*], and Yanpeng Qi[1,3*]

[1]School of Physical Science and Technology, ShanghaiTech University, Shanghai 201210, China

[2]Department of Physics and Beijing Key Laboratory of Opto-electronic Functional Materials & Micro-nano Devices, Renmin University of China, Beijing 100872, China

[3]ShanghaiTech Laboratory for Topological Physics, ShanghaiTech University, Shanghai 201210, China

[#] These authors contributed to this work equally.

[*] Correspondence should be addressed to Y.Q. (qiyp@shanghaitech.edu.cn) or K.L. (kliu@ruc.edu.cn) or H.C.L. (hlei@ruc.edu.cn)

**ABSTRACT:**

**The recent discovery of superconductivity up to 32 K in the pressurized $MoB_2$ reignites the interests in exploring high-$T_c$ superconductors in transition-metal diborides. Inspired by that work, we turn our attention to the 5$d$ transition-metal diborides. Here we systematically investigate the responses of both structural and physical properties of $WB_2$ and $ReB_2$ to external pressure, which possess different types of boron layers. Similar to $MoB_2$, the pressure-induced superconductivity was also observed in $WB_2$ above 60 GPa with a maximum $T_c$ of 15 K at 100 GPa, while no superconductivity was detected in $ReB_2$ in this pressure range. Interestingly, the structures at ambient pressure for both $WB_2$ and $ReB_2$ persist to high pressure without structural phase transitions. Theoretical calculations suggest that the ratio of flat boron layers in this class of transition-metal diborides may be crucial for the appearance of high $T_c$. The combined theoretical and experimental results highlight the effect of geometry of boron layers on superconductivity and shed light on the exploration of novel high-$T_c$ superconductors in borides.**

## INTRODUCTION

The key structural unit in materials often has a significant impact on the physical properties. A paradigm is the [$CuO_2$] layers in cuprate superconductors [1]. Although the superconducting critical temperature $T_c$ of cuprate superconductors depends on the specific materials, it has been well established that the existence of two-dimensional (2D) [$CuO_2$] conducting layers modulated by neighboring charge reservoir layers is essentially important for the high-$T_c$ superconductivity. Similar to the [$CuO_2$] layers in cuprate superconductors, the [FeAs] layer is the key structural element in the FeAs-based superconductors [2]. Meanwhile, for the widely studied metal-superhydride under ultra-high pressure [3-9], the hydrogen clathrate structure is also considered to be important to the high-$T_c$ superconductivity. With the fine tuning of the geometries of these key structural units, the $T_c$ can be effectively modulated. Empirically, the extent of distortion of $FeAs_4$ tetrahedra in FeAs-based superconductors is crucial to $T_c$, magnetism, and correlation strength [10-12]. Therefore, identifying the key structural unit and tuning its geometry are very important to explore novel superconductors and deepen the understanding of relationship between superconductivity and crystal structure.

Beyond the cuprate and iron-based superconductors, $MgB_2$ is another template superconductor with the key structural unit of boron layers. The superconducting $T_c$ of 39 K in $MgB_2$ has been considered to approach the McMillan limit [13]. Theoretical calculation indicates that the superconductivity in $MgB_2$ is intimately related to the vibration of B atoms and the B-B bonding [14-17]. Motivated by these results, we carried out a high-pressure study on the transition-metal diborides and surprisingly we discovered that the application of high pressure drives the $MoB_2$ to a superconductor with $T_c$ as high as 32 K, which is close to that of $MgB_2$ and the highest among the transition-metal diborides [18]. The superconducting $T_c$ of $MoB_2$ firstly emerges at 30 GPa and then increases rapidly until 70 GPa, where it undergoes a structural transition to the $MgB_2$-type structure. After this structural transition, $T_c$ varies gently with the applied pressure. For the superconductivity of $MoB_2$ in the $MgB_2$-type structure at high pressure, theoretical calculations have indicated that it still belongs to the conventional BCS-type [18, 19]. Nevertheless, the origin of the emergent superconductivity before this structural transition has not been resolved.

In order to have a comprehensive understanding of physical properties of transition-metal diborides and the role of B in superconductivity, in this work we study the high-pressure responses of $5d$ transition-metal diborides $WB_2$ and $ReB_2$. At ambient pressure, both $WB_2$ and $ReB_2$ have similar triangular metal layers, (Figs. S1a and S1b) [20] [21] however, their local arrangements of B atomic layers are quite different. There are two different kinds of B layers in WB$_2$ (Fig. 1): one forms a nearly planar quasi-2D honeycomb lattice similar to that in MgB$_2$; the other builds a buckled honeycomb network. At ambient pressure, the ratios between the planar and buckled B layers in WB$_2$ is 1 : 1, similar to MoB$_2$. On the other hand, ReB$_2$ only has buckled honeycomb B layers without planar ones. Because of different structural configurations of B layers, these two 5$d$ transition-metal diborides provide a good material platform to make a comparative study with MgB$_2$ and MoB$_2$.

With applying pressure, it is found that WB$_2$ exhibits superconductivity at around 60 GPa, reaching a maximum $T_c$ of 15 K at around 100 GPa, while superconductivity is still absent in ReB$_2$ in this pressure range. Synchrotron x-ray diffraction measurements demonstrate that their ambient-pressure structure persists in the high-pressure region without structural phase transition. The first-principle calculations indicate that pressure tends to drive more flat B layers in WB$_2$ and the structure with more flat B layers will favor a higher $T_c$.

**EXPERIMENTAL AND COMPUTATIONAL METHODS**

**Sample synthesis, structural and composition characterizations.** Single crystals of WB$_2$ and ReB$_2$ were grown by Al flux. Mo/Re (99.5 %), B (99.9 %) and Al (99.99 %) with a molar ratio of Mo/Re : B : Al = 1 : 2.5 : 73.3 were put into an alumina crucible. The mixture was heated up to 1773 K in a high-purity argon atmosphere. Then it was cooled down to 1173 K. The single crystals were separated from the Al flux using sodium hydroxide solution. The powder x-ray diffraction (XRD) pattern was performed using a Bruker D8 x-ray diffractometer with Cu $K_\alpha$ radiation ($\lambda$ = 1.5418 Å) at room temperature.

**Experimental details of high-pressure measurements.** High pressure resistivity measurements were performed in a nonmagnetic diamond anvil cell as described elsewhere [22-24]. A cubic BN/epoxy mixture layer was inserted between BeCu gaskets and electrical leads. Electrical resistivity was measured using the dc current in van der Pauw technique in Physical Property Measurement System (Dynacool, Quantum Design). Pressure was measured using the ruby scale by measuring the luminescence from small chips of ruby placed in contact with the sample.

*In situ* high pressure XRD measurements were performed at the beamline 15U at Shanghai Synchrotron Radiation Facility ($\lambda$ = 0.6199 Å). Symmetric diamond anvil cell (DAC) with anvil culet sizes of 200 μm as well as Re gaskets were used. Mineral oil was used as PTM and pressure was determined by the ruby luminescence method[25]. CeO$_2$ was used to calibrate the sample-detector distance and the orientation parameters of the detector. The two-dimensional diffraction images were analyzed using the FIT2D

[26] program. Rietveld refinements on crystal structures under high pressure were performed by General Structure Analysis System (GSAS) and graphical user interface EXPGUI package [27, 28].

**Theoretical calculations.** The electronic structures, phonon spectra, and electron-phonon couplings (EPC) of $WB_2$ in the $WB_2$-type, $MgB_2$-type, $MoB_2$-type, $ReB_2$-type structures and $ReB_2$ in the $ReB_2$-type structure were studied based on the density functional theory (DFT) [29, 30] and density functional perturbation theory (DFPT) [31, 32] calculations as implemented in the Quantum ESPRESSO (QE) package [33]. The interactions between electrons and nuclei were described by the norm-conserving pseudopotentials [34]. For the exchange-correlation functional, the generalized gradient approximation of Perdew-Burke-Ernzerhof type was adopted [35]. The kinetic energy cutoff of the plane-wave basis was set to be 80 Ry. For different structure phases, we firstly performed the calculations in their conventional cells. A 24×24×(24/N) ***k***-point mesh was used for the sampling of Brillouin zone (BZ), where N is the number of the B layer in their conventional cells. The Gaussian smearing method with a width of 0.004 Ry was employed for the Fermi surface broadening. In the structural optimization, both lattice constants and internal atomic positions were fully relaxed until the forces on atoms were smaller than 0.0002 Ry/Bohr.

The superconductivities of $WB_2$ and $ReB_2$ in different structural phases and different pressure conditions were studied based on the EPC theory as implemented in the EPW package [36], which uses the maximally localized Wannier functions (MLWFs) [37] and interfaces with QE [33]. Our adopted coarse and dense grids of ***k***-mesh and ***q***-mesh for different structural phases are shown in Table 1.

Table 1. The coarse and dense grids of ***k***-mesh and ***q***-mesh used for the EPC calculations of $WB_2$ and $ReB_2$.

|  | $WB_2$-type | $MgB_2$-type | $MoB_2$-type | $ReB_2$-type |
|---|---|---|---|---|
| coarse ***k***-mesh | 6×6×4 | 6×6×6 | 8×8×8 | 6×6×4 |
| coarse ***q***-mesh | 6×6×2 | 6×6×6 | 4×4×4 | 6×6×4 |
| dense ***k***-mesh | 48×48×16 | 48×48×48 | 48×48×48 | 48×48×32 |
| dense ***q***-mesh | 24×24×8 | 24×24×24 | 16×16×16 | 24×24×16 |

Specially, the EPC calculations of $MoB_2$-type structure were performed in its rhombohedral primitive cell. The EPC constant $\lambda$ can be calculated either by the summation of the EPC strength $\lambda_{qv}$ in the whole BZ for all phonon modes or by the integral of the Eliashberg spectral function $\alpha^2F(\omega)$ [38] as following,

$$\lambda = \sum_{qv} \lambda_{qv} = 2\int \frac{\alpha^2 F(\omega)}{\omega} d\omega.$$

The Eliashberg spectral function $\alpha^2F(\omega)$ is defined as [38],

$$\alpha^2F(\omega) = \frac{1}{2\pi N(\varepsilon_F)}\sum_{q\nu}\delta(\omega-\omega_{q\nu})\frac{\gamma_{q\nu}}{\hbar\omega_{q\nu}},$$

where $N(\varepsilon_F)$ is the density of states at the Fermi level $\varepsilon_F$, $\omega_{q\nu}$ is the frequency of the $\nu$-th phonon mode at the wave vector $\boldsymbol{q}$, and $\gamma_{q\nu}$ is the phonon linewidth. The superconducting transition temperature $T_c$ can be calculated by substituting the EPC constant $\lambda$ into the McMillan-Allen-Dynes formula [39, 40],

$$T_c = \frac{\omega_{log}}{1.2}\exp[\frac{-1.04(1+\lambda)}{\lambda(1-0.62\mu^*)-\mu^*}],$$

where $\omega_{log}$ is the logarithmic average frequency defined as [39, 40]

$$\omega_{log} = \exp[\frac{2}{\lambda}\int\frac{d\omega}{\omega}\alpha^2F(\omega)\ln(\omega)],$$

and $\mu^*$ is the effective screened Coulomb repulsion constant that was set to an empirical value of 0.1 [41, 42].

**RESULTS AND DISCUSSION**

At ambient pressure, both WB$_2$ and ReB$_2$ show the metallic behavior (Figs. S1c and S1d). We measured electrical resistivity $\rho(T)$ for both samples at various pressures. Figure 2a shows the typical $\rho(T)$ curves of WB$_2$ single crystal for pressure up to 120 GPa. In the low-pressure range, the room-temperature resistivity drops dramatically, while it changes mildly for $P > 20$ GPa. When the pressure increases to 50 GPa, a small broad drop of $\rho$ is observed at the lowest measuring temperature ($T_{min} = 1.8$ K), as shown in Fig. 2b. With further increasing pressure, the superconducting $T_c$ goes up rapidly and zero resistivity is achieved at low temperature for $P > 98$ GPa, indicating the emergence of superconductivity. Additional pressure increases have only a weak effect on the onset superconducting transition temperature, but the transition becomes much sharper. The maximum $T_c$ of 15 K is attained at $P = 100$ GPa. Beyond this pressure $T_c$ decreases slowly, but superconductivity persists up to the highest experimental pressure ($P \approx 120$ GPa).

The measurements on different samples of WB$_2$ for three independent runs provide the consistent and reproducible results, confirming the intrinsic superconductivity under pressure[43] (Fig. S2). The pressure dependences of the resistivity in the normal state at $T = 300$ K and of the critical temperature of superconductivity for WB$_2$ are summarized in Fig. 2c. No trace of superconductivity is observed down to 1.8 K at low pressure range. Superconductivity suddenly appears at and above 65 GPa, with $T_c$ rapidly rising to 15 K. At the same time, the room temperature resistivity drops monotonically.

To gain insights into the superconducting transition, we conducted resistivity measurements in the vicinity of $T_c$ under various external magnetic fields. As can be seen in Fig. 2d, the resistivity drops under $P$ = 121.3 GPa are gradually suppressed with increasing field. Such behavior further confirms that the sharp decrease of $\rho(T)$ should originate from a superconducting transition. The derived upper critical field $\mu_0H_{c2}(T)$ as a function of temperature $T$ can be fitted well using the empirical Ginzburg-Landau (GL) formula (Fig. 2e) $\mu_0H_{c2}(T) = \mu_0H_{c2}(0)(1 - t^2)/(1 + t^2)$, where $t = T/T_{c0}$ is the reduced temperature with zero-field superconducting transition temperature $T_{c0}$. The fitted zero-temperature upper critical field $\mu_0H_{c2}(0)$ of WB$_2$ from the 90% $\rho_n$ criterion reaches 2.00(1) T at 121.3 GPa, which yields a Ginzburg–Landau coherence length $\xi_{GL}(0)$ of 12.81(1) nm. It is also worth noting that the estimated value of $\mu_0H_{c2}(0)$ is well below the Pauli-Clogston limit (= $1.865T_{c0}$ = 27.12 T). Fig. 2f shows temperature dependence of resistivity of ReB$_2$ single crystal at various pressure. Compared with WB$_2$, no superconductivity in ReB$_2$ was observed down to 1.8 K in this pressure range.

At ambient pressure, both WB$_2$ and ReB$_2$ crystallize in a hexagonal structure with space group $P6_3/mmc$. Since a pressure-induced structural phase transition from the CaSi$_2$ structure to the MgB$_2$ structure above 70 GPa has been observed in MoB$_2$ single crystal [18], in order to check the structure stability under pressure, *in situ* XRD measurements on pulverized WB$_2$ as well as ReB$_2$ crystals were also performed under various pressures at room temperature. Fig. 3a exhibits the high-pressure synchrotron XRD patterns of WB$_2$ measured up to 112.4 GPa. All the diffraction peaks in the low-pressure range can be indexed well to a hexagonal primitive cell of WB$_2$ (space group $P6_3/mmc$), and both *a*-axis and *c*-axis lattice constants decrease with increasing pressure (Fig. S4a). The representative refinements at 1 atm and selected pressures are shown in Supplemental Information (Fig. S3). We note that *c*/*a* increases with pressure and has a slope change at ~ 68 GPa (Fig. 3b). It should be noted that the superconductivity is observed beyond this pressure. This slope change might be an indication of a change in the microstructural features of the sample at high pressure. We also investigated the structural evolution of ReB$_2$ under pressure (Figs. 3c, 3d and Fig. S4b). High-pressure XRD patterns demonstrate the robustness of ambient structure and a lack of structural phase transition in ReB$_2$.

We observed pressure-induced superconductivity in WB$_2$ and MoB$_2$ [18] but not in ReB$_2$ with the same experimental conditions. The major difference between WB$_2$ and ReB$_2$ is the local arrangements of B atomic layers, i.e., the different numbers of flat and buckled B layers. It seems that the more flat B layers the transition metal diboride like MgB$_2$ or high-pressure phase MoB$_2$ has, the higher $T_c$ there is. Thus, in order to reveal the possible origin of such trend and understand the distinct superconducting behaviors of WB$_2$ and ReB$_2$, we first calculated the enthalpies of WB$_2$ with different hypothetic structures in the pressure range of 0 to 150 GPa. As can be seen from Fig. 4a, the ReB$_2$-type structure has the lowest enthalpy below 12 GPa. Between 12 and 127 GPa, the MoB$_2$-type structure is energetically favored. Beyond 127 GPa, the MgB$_2$-type structure becomes the most stable one. In the whole pressure range, the real structure of WB$_2$ seems to be a metastable phase. Nevertheless, as labeled by the percentage numbers in Fig. 4a, the ratio of flat B layers in the lowest-energy structural phases becomes larger and larger with the increasing pressure, indicating that the pressure tends to drive the buckled B layers to become the flat ones. The evolutions of lattice parameters with applied pressure are shown in Figs. 4b and 4c. The in-plane lattice constants $a$ of all structural phases decreases with the pressure, but the ratios of interlayer distance $d$ to in-plane lattice constant $a$ evolve differently. Only the $d/a$ ratio in MgB$_2$-type structure with 100% flat B layers reduces with pressure and those of other structures enlarge slightly. The sudden flattening of some B layers in real samples under high pressure may induce a discontinuity in the $d/a$ ratio, as evidenced by the kink at 68.8 GPa in our experimental data (Fig. 3b).

The structural change in B layers could influence the electronic structures. As to the calculated electronic density of states at the Fermi level $N(\varepsilon_F)$ (Fig. 4d), the more the flat B layers are in the structures, the higher $N(\varepsilon_F)$ will the structures have. Correspondingly, the calculated superconducting $T_c$'s for different structural types of WB$_2$ are displayed in Fig. 4e. The ReB$_2$-type structure with 0% flat B layers is not superconducting at 0, 50, and 100 GPa. The WB$_2$-type and MoB$_2$-type structures with 50% flat B layers show superconductivity below 5 K. In comparison, the $T_c$ of MgB$_2$-type structure with 100% flat B layers is around 25 K at 100 and 150 GPa. According to Fig. 2, the measured $T_c$ of WB$_2$ sample at 100 GPa is ~ 15 K, which suggests that the real WB$_2$ sample may contain a ratio of flat B layer between 50% and 100% under high pressure. On the other hand, our calculated $T_c$'s for ReB$_2$ in its ReB$_2$-type structure with 0% flat B layers (Fig. 1) always keep 0.0 K among 0 ~ 100 GPa (Fig. 4e), consisting with our experimental observation of the absence of superconductivity in ReB$_2$ (Fig. 2f).

To demonstrate the relationship between superconductivity and structural unit more clearly, the calculated superconducting $T_c$, the electronic density of states $N(\varepsilon_F)$, and the ratio of flat B layers for WB$_2$ in above structural types are further summarized in Fig. 5. As shown in Fig. 5a, there is a linear relationship between $\ln(T_c/\omega_D)$ and $-1/N(\varepsilon_F)$. This indicates that the $T_c$ does have a positive correlation with the $N(\varepsilon_F)$, which fulfills the BCS mechanism: $k_B T_c/\hbar\omega_D = 1.13\exp[-1/N(\varepsilon_F)V]$ [44, 45], where $k_B$ is the Boltzman constant, $\hbar$ is the Planck constant, and $V$ is the paring potential. From Fig. 5b, the structures with larger ratio of flat B layers own higher density of states $N(\varepsilon_F)$. Thus, we can draw the conclusion that the structures with more flat B layers will have a higher $T_c$. This is in accordance with our experimental hints on WB$_2$, whose kink of *c*/*a* ratio (flattening of some B layers) and superconductivity emerge at the same pressure (Figs. 2f and 3b). It also agrees with our previous observations on MoB$_2$, which exhibits a superconducting $T_c$ of 32 K when MoB$_2$ transits to the MgB$_2$-type structure with 100% flat B layers under high pressure [18]. Besides the effect of pressure, the ratio of flat B layers in real WB$_2$ sample may exceed 50% due to the stacking faults in the synthesis process. Our calculations indicate that the extra flat B layers will induce multiple kinds of lattice distortions at low pressure but recover to flat ones at high pressure, which influence the $N(\varepsilon_F)$ and hence the emergence of superconductivity (Fig. S5). As for the absent superconductivity in pressed ReB$_2$, it's probably because the ratio of flat B layers in ReB$_2$ always keeps or approaches 0% among the whole investigated pressure region. Combined with our previous study on MoB$_2$ [18], for this class of transition-metal diborides, both the stacking faults of crystals or the application of pressure may help to enhance the superconductivity by the flattening of B layers.

**CONCLUSION**

In summary, we have measured the resistivity transitions of WB$_2$ and ReB$_2$ crystals under pressure up to 120 GPa, which possess different types of boron layers and diverse stacking orders. Pressure-induced superconductivity in WB$_2$ begins near 60 GPa with a maximum superconducting $T_c$ of 15 K at 100 GPa, while no superconductivity was detected in ReB$_2$ in this pressure range. According to our x-ray analysis, the structures at ambient pressure for both WB$_2$ and ReB$_2$ persist to high pressure without structural phase transitions. Our first-principles calculations indicate that the ratio of flat boron layers in this class of transition-metal diborides may be crucial for the appearance of high $T_c$. Our study highlights the effect of geometry of boron layers on superconductivity and paves a novel way to induce superconductivity under pressure.

ACKNOWLEDGMENT

This work was supported by the National Key R&D Program of China (Grant No. 2018YFA0704300, 2018YFE0202600 and 2017YFA0302903), the National Natural Science Foundation of China (Grant No. U1932217, 11974246, 12004252, 12174443, 11822412, 11774423 and 11774424), the Natural Science Foundation of Shanghai (Grant No. 19ZR1477300), the Science and Technology Commission of Shanghai Municipality (19JC1413900), the Beijing Natural Science Foundation (Grant No. Z200005), and the Fundamental Research Funds for the Central Universities and Research Funds of Renmin University of China (RUC) (Grant No. 19XNLG17). The authors thank the support from Analytical Instrumentation Center (# SPST-AIC10112914), SPST, ShanghaiTech University. The authors thank the staffs from BL15U1 at Shanghai Synchrotron Radiation Facility for assistance during data collection. Computational resources were provided by the Physical Laboratory of High Performance Computing at Renmin University of China and Beijing Super Cloud Computing Center.

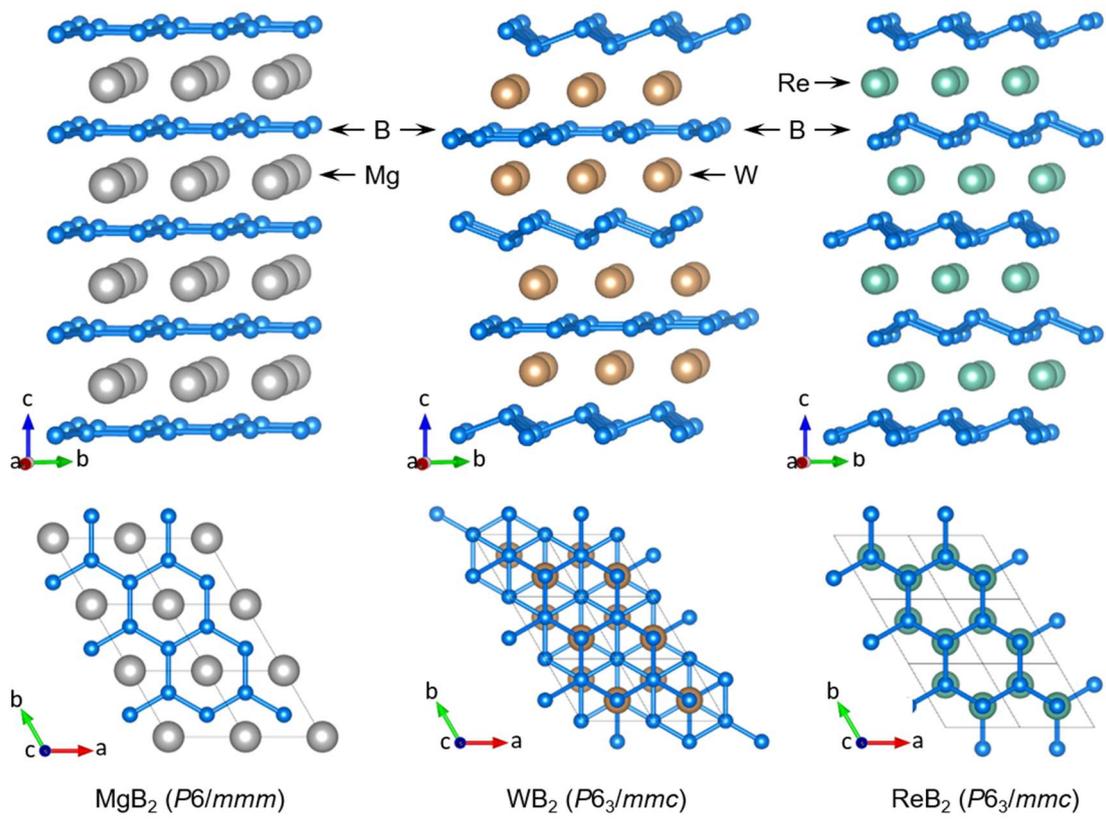

Figure 1. Crystal structures of MgB$_2$ (*P*6/*mmm*), WB$_2$ (*P*6$_3$/*mmc*) and ReB$_2$ (*P*6$_3$/*mmc*).

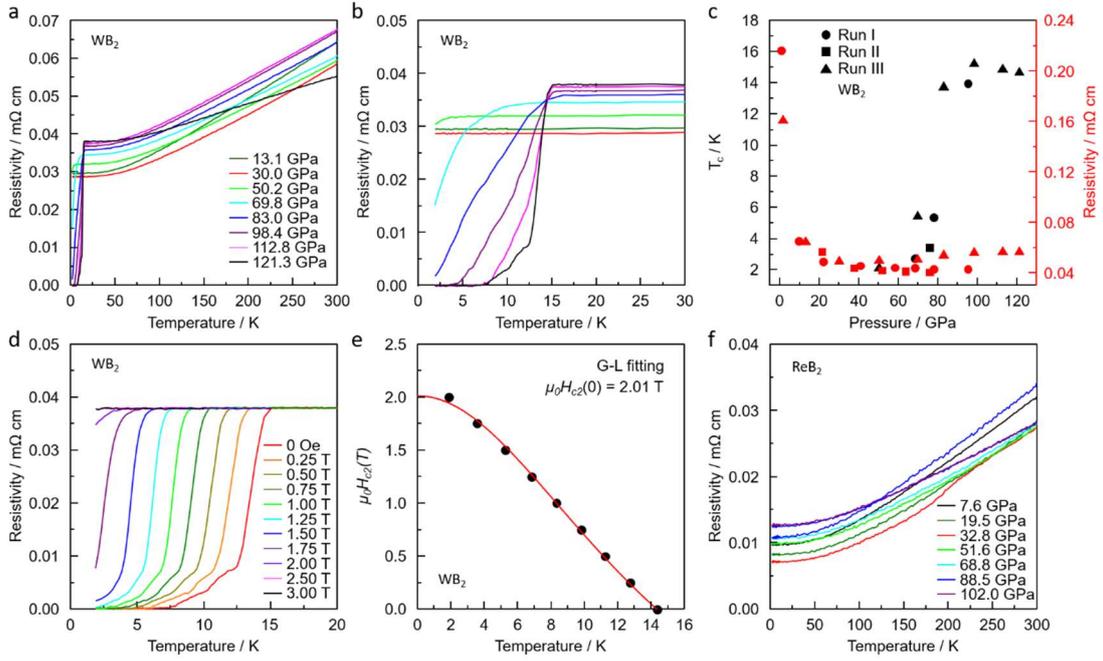

Figure 2. (a) Electrical resistivity of $WB_2$ as a function of temperature under different pressures; (b) Temperature-dependent resistivity of $WB_2$ in the vicinity of the superconducting transition; (c) Pressure dependence of the superconducting transition temperatures $T_c$s and the resistivities at 300 K for $WB_2$ up to 121.3 GPa in different runs; (d) Temperature dependence of resistivity under different magnetic fields for $WB_2$ at 121.3 GPa; (e) Temperature dependence of upper critical field for $WB_2$ at 121.3 GPa. Here, $T_c$ is determined as the 90% drop of the normal state resistivity. The red solid lines represent the fits based on the GL formula; (f) Electrical resistivity of $ReB_2$ as a function of temperature under various pressures.

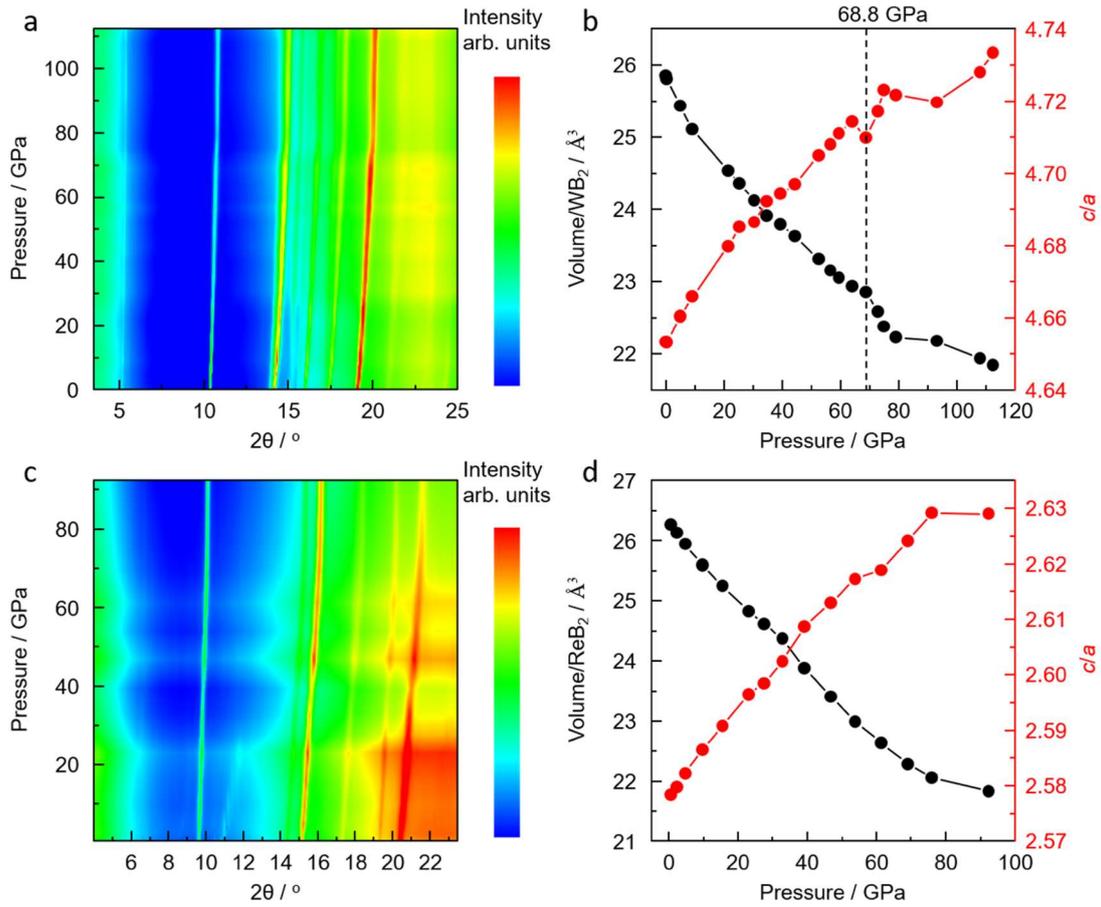

Figure 3. Contour color plot of pressure dependence of the XRD pattern of (a) $WB_2$ up to 112.4 GPa and (c) $ReB_2$ up to 92.4 GPa, respectively. Pressure-dependence of lattice volume and *c/a* ratio for (b) $WB_2$ in *P6_3/mmc* structure and (d) $ReB_2$ in *P6_3/mmc* structure, respectively.

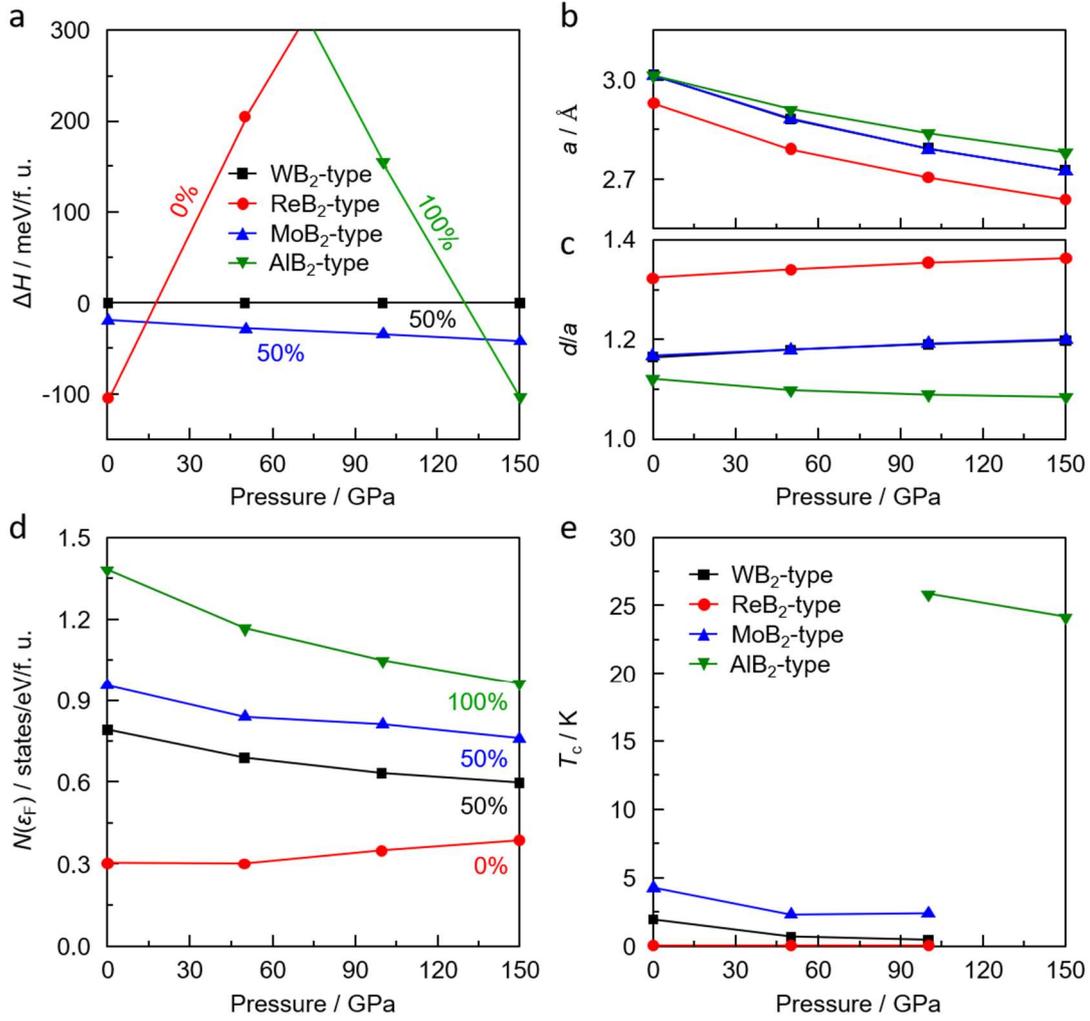

Figure 4. (a) The calculated enthalpy differences $\Delta H$ of $WB_2$ with different hypothetic structures in comparison with the real $WB_2$-type structure. The numbers near each line label the ratio of planar B layers in the corresponding structural phases. (b) The relaxed lattice constant $a$ and (c) the ratio of $d/a$ per formula unit. (d) The calculated electronic density of states at the Fermi level $N(\varepsilon_F)$ and (e) the calculated superconducting $T_c$ based on EPC theory for $WB_2$ in different structural phases.

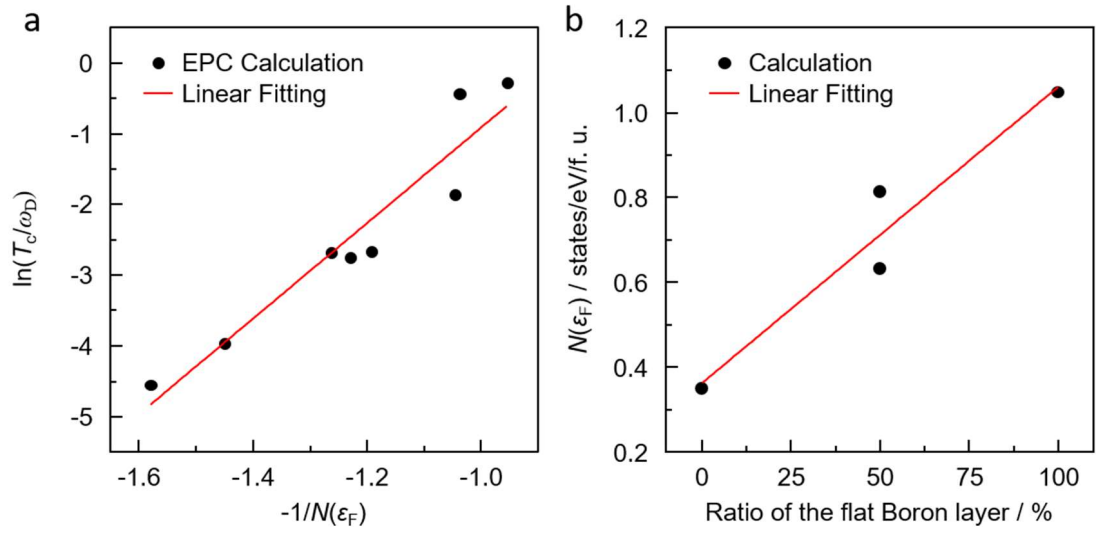

Figure 5. (a) The relation between our calculated $\ln(T_c/\omega_D)$ based on the EPC theory and $-1/N(\varepsilon_F)$. The data (black dots) are gathered from the non-zero values in Fig. 5(e). (b) The relation between the electronic density of states at the Fermi level $N(\varepsilon_F)$ under 100 GPa and the ratio of flat boron layers in corresponding structural phases.